\documentclass{PoS}

\title{Measurement of the Z boson plus two b-jets cross section in CMS with 100
pb$^{-1}$}

\ShortTitle{Measurement of the Z boson plus two b-jets cross section in CMS with 100 pb$^{-1}$}

\author{\speaker{Aruna Kumar NAYAK}%
        \thanks{On behalf of CMS Collaboration}\\
        Tata Institute of Fundamental Research, Mumbai\\
        E-mail: \email{aruna.nayak@cern.ch}}

\abstract{The cross section for production of Z bosons with two associated b-jets, and Z decaying to leptons, can be measured at the LHC with about 100 pb$^{-1}$ of data. We use simulated data to study possible strategies for an early measurement of this process with the CMS detector. The rate and kinematic properties of this final state needs to be well understood because it constitutes a large fraction of the total backgrounds to several of the Higgs discovery channels at the LHC.}

\FullConference{Physics at LHC 2008\\
        29 September - October 4, 2008\\
        Split, Croatia}

\begin{document}

\section{Introduction}
At the LHC, $Z/\gamma^{*}$ can  be produced in association with two b-quarks,
$gg/q\bar{q}\rightarrow b\bar{b}Z$, with a significant cross section~\cite{Campbell:2000bg}, \cite{Campbell:2003hd},
\cite{Maltoni:2005wd}. The measurement of this process is an important test of QCD calculations and also can help to reduce the uncertainty of the Super Symmetry $b\bar{b}H$ production cross section calculation \cite{Campbell:2004},  since the two processes are produced via similar partonic processes at the initial state. This process is also a background to  Higgs boson discovery channels at the LHC, like Standard Model $H \to ZZ \to 4 \ell$ and associated production of SUSY Higgs boson $bb\varphi,~\varphi \to \tau \tau~ (\mu \mu)$ ($\varphi=h,~H,~A$).
In the following, the possibility of observing 
and measuring the cross section of the $b\bar{b}Z$ production at the LHC, using 100~pb$^{-1}$ of 
early data at CMS experiment, has been described which was studied in \cite{PAS_EWK_08_001}. A related, but different cross section for 
$p\bar{p}\rightarrow Z + b$-jet has previously been measured at the Tevatron, both by D0 and CDF 
\cite{Abulencia:2006ce}, \cite{Abazov:2004zd}. 

\section{Monte Carlo Signal and Background samples}

The signal events $\ell\ell b\bar{b}$ ($Zbb$ ) were generated at parton level using the leading order (LO) CompHEP generator 
~\cite{Pukhov:1999gg} and hadronized with PYTHIA~\cite{Sjostrand:2003wg} with the following generator-level cuts : $p_{t}^{b}>$10 GeV/$c$, $|\eta^{b}|<$10, $m_{\ell\ell}>$ 40 GeV/$c^2$ 
and $|\eta^{\ell}|<$2.5. The next-to-leading order cross-section $\sigma(\ell \ell b \bar{b})$ = 45.9~pb ($\ell$=e, $\mu$, $\tau$) 
was calculated with the MCFM program ~\cite{Campbell:2002tg} applying the same generator-level cuts. The CTEQ6M parton density functions
(pdf) and scale $\mu_{R}=\mu_{F}=M_{Z}$ were used. 
The backgrounds considered for this process were Drell-Yan $Z/\gamma^{*} \to \ell \ell ~ (\ell = e, \mu, \tau)$ production in
association with two or more light-quark and gluon jets ($Z$+jets), $\ell\ell c\bar{c}$+jets ($Zcc$+jets) and $t\bar{t}$+jets. 
Background samples were generated at the leading order (LO) with the ALPGEN generator ~\cite{Mangano:2002ea}.
The $t\bar{t}$+jets ALPGEN events were normalized to the NLO inclusive $t\bar{t}$ cross section 840~pb~\cite{Beneke:2000hk}.
The $Zcc$+jets events were normalized to the NLO MCFM cross section of $\ell \ell c \bar{c}$, 13.29~pb, applying
the ALPGEN production cuts: $p_{T}^{c}>$20 GeV/$c$, $|\eta^{c}|<$5, $m_{\ell\ell}>$ 40 GeV/$c^2$. The
same pdf and scale settings as for $\ell \ell b \bar{b}$ process were used.  
The $Z$+jets events were normalized to NLO MCFM cross section of $\ell \ell$~+~2~jets, 714~pb, applying
the ALPGEN production cuts: $p_{T}^{j}>$20 GeV/$c$, $|\eta^{j}|<$5, $m_{\ell\ell}>$ 40 GeV/$c^2$. 
Signal and background samples were passed through the full simulation and reconstruction chain of CMS, 
under an imperfect calibration and alignment configuration assumed to be typical of the first 
100~pb$^{-1}$ of data at CMS.

\begin{figure}[hbtp]
\begin{center}
\includegraphics[width=0.35\textwidth]{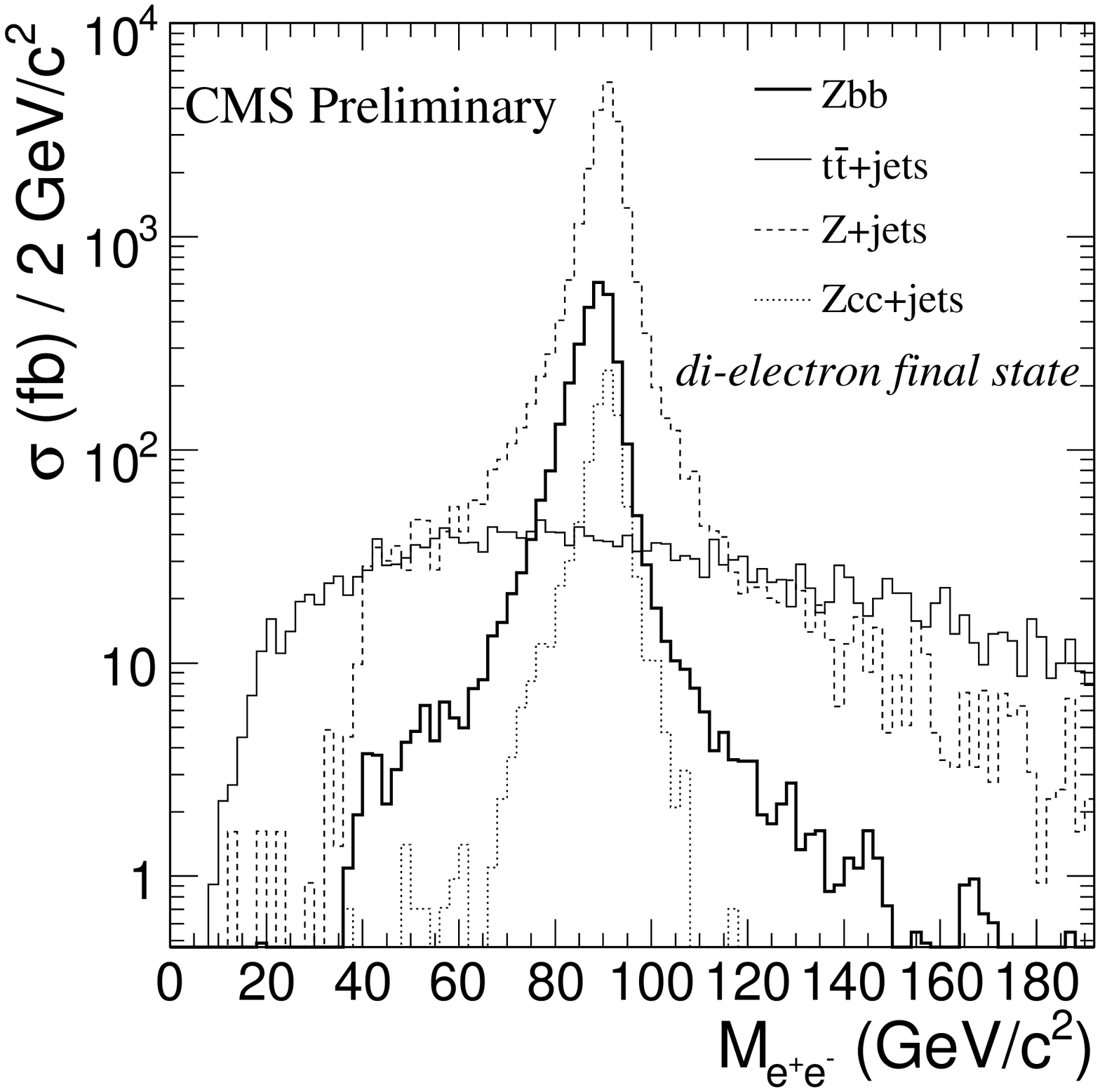}\includegraphics[width=0.35\textwidth]{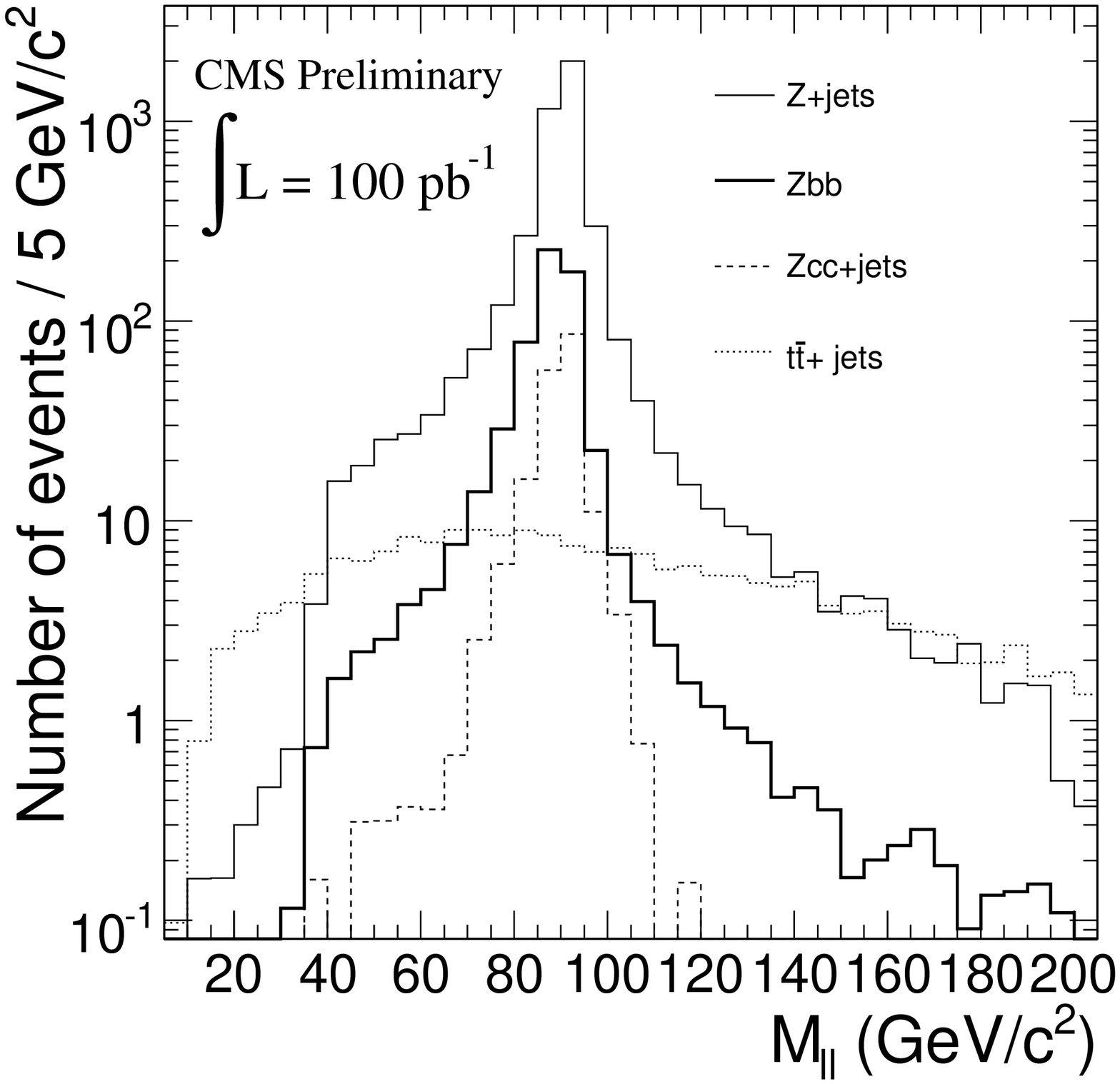}
\caption{Left : Di-lepton invariant mass distribution for events passing lepton and jet selections as described in text. Right : Di-lepton invariant mass for events after all selections except b-tagging.}
\label{primdiLepMass}
\end{center}
\end{figure}

\section{Event Selection}
\begin{figure}[hbtp]
\begin{center}
\includegraphics[width=0.39\textwidth]{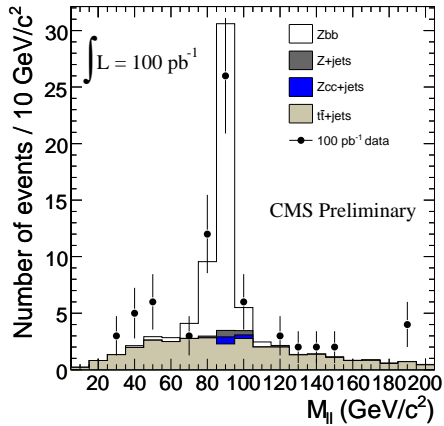}
\caption{Di-lepton invariant mass for events after all selections. The events are normalized to 100~pb$^{-1}$ of data. The dots are result of one random MC experiment with 100~pb$^{-1}$ of data. Nearly 46 events are expected in the mass range 75-105 GeV with S/B $\sim$ 3.6}
\label{diLepMassFinal}
\end{center}
\end{figure}

The events were selected by the CMS Level-1 (L1) and High-Level (HLT) single isolated electron and muon triggers designed for the low luminosity period (L=10$^{32}$cm$^{-2}$s$^{-1}$). In offline, Events with at least two isolated and oppositely charged electrons or muons, with $p_{T}^{e,\mu}$ greater than 20 GeV, $|\eta^{e}|$ $<$ 2.5 and $|\eta^{\mu}|$ $<$ 2.0,  were selected to form $Z$ boson candidate. Events are selected with two or more jets, reconstructed using iterative cone algorithm of cone size 0.5,  with corrected jet $E_{T}$ greater than 30 GeV and jet $|\eta| < 2.4$. The $\eta$ cut ensured good quality b-tagging. The Figure ~\ref{primdiLepMass} (Left) 
shows the distribution of di-lepton invariant mass for the electron final state events with at least two jets passing lepton and jet selections. The b-tagging is an important and effective tool to ensure the purity of $Zbb$ events and reduce the $Z+jets$ and  $Zcc$+jets backgrounds since the b-tag discriminator values for light quarks, gluon and c-quark jets tends to be lower compared to that of the b-quark jets. The events were double b-tagged using 'track counting' b-tagging, 
which uses  the 3D track impact parameter significance of the 3rd highest significance track in the jet as the b-tagging discriminator. A b-tag discriminator value of 2.5 has been used in the analysis which ensured high purity of b-tagged jets. The possibility of applying selections on the amount of missing transverse energy in the event, $E_{T}^{miss}$, is exploited to suppress $t\bar{t}$+jets background events since $t\bar{t}$+jets events contain neutrinos from $W \rightarrow \ell \nu$ decays and has higher $E_{T}^{miss}$ compared to $Zbb$ events. A selection cut on $E_{T}^{miss}$ less than 50 GeV is applied on $E_{T}^{miss}$, where $E_{T}^{miss}$ is reconstructed using calorimeter tower information and calibrated using jet energy corrections and muon corrections. The dilepton invariant mass 
distributions of the signal and various background events passing all selection criteria are shown in 
Figure ~\ref{diLepMassFinal}. The events are scaled to 100~pb$^{-1}$ of data.

\section {Background Estimation and Systematic Uncertainties}

The contribution of the $t\bar{t}$+jets background events in the signal region of the di-lepton invariant mass distribution 
can be estimated from the data, by extrapolating sidebands in the di-lepton invariant mass spectrum.
The following method is used to estimate the $t\bar{t}$+jets background using the relation  
$N_{Z}(t\bar{t}) = (\epsilon_{Z}(t\bar{t})/\epsilon_{noZ}(t\bar{t})) \times N_{noZ}(t\bar{t})$, where $N_{Z}(t\bar{t})$=8 is the expected number of 
$t\bar{t}$+jets events in the signal region within the $m_{\ell \ell}$ mass window of 75-105 GeV/$c^2$, 
$N_{noZ}(t\bar{t})$=27 is the measured number of $t\bar{t}$ events outside the signal region, $\epsilon_{Z}(t\bar{t})$ is the 
selection efficiency for $t\bar{t}$+jets events in signal region, $\epsilon_{noZ}(t\bar{t}))$ is the selection efficiency for 
$t\bar{t}$+jets events outside the signal region. $\Delta N_{Z}(t\bar{t})$ is the uncertainty of the expected number of 
$t\bar{t}$+jets events in the signal region given by  $\Delta N_{Z}(t\bar{t})/N_{Z}(t\bar{t}) = 1/\surd N_{noZ}(t\bar{t})$. The uncertainty on the ratio $\epsilon_{Z}(t\bar{t})/\epsilon_{noZ}(t\bar{t})$ 
is negligible in comparison to the statistical uncertainty of the number of events outside the signal region. 

The following equations are used to determine the number of $Zbb$ events ($N_{Zbb}$) and it's uncertainty $\Delta N_{Zbb}$ for 
100~pb$^{-1}$, after all selections except double b-tagging. 

\begin{eqnarray}
  N_{Z}^{before~b-tag}~=~N_{Zjj}~+~N_{Zcc}~+~N_{Zbb} \\
  N_{Z}^{after~b-tag}~=~\varepsilon _{\ell} \times N_{Zjj}~+~\varepsilon _{c} \times N_{Zcc}~+~\varepsilon _{b} \times N_{Zbb}
\end{eqnarray}
where,  $N_{Z}^{before~b-tag}$ = 4644 is the measured number of $Z/\gamma^{*} \to \ell \ell$ events in the di-lepton mass window between 75 and 105 GeV/$c^2$ after all selections have been applied except double b-tagging. It receives negligible contribution (less than 1\%) from the $t\bar{t}$ background, as one can see from Figure~\ref{primdiLepMass} (Right). 
$N_{Z}^{after~b-tag}$ = 38 is the measured number of $Z/\gamma^{*} \to \ell \ell$ events after all selections including double b-tagging. It is defined after subtraction of the $t\bar{t}$ background described above.
$N_{Zjj}$ is the unknown number of $\ell \ell$+jets (jet=u,d,s,g) events before double b-tagging.
$N_{Zcc}$ is the unknown number of $\ell \ell c \bar{c}$ events before double b-tagging.
$N_{Zbb}$ is the unknown number of $\ell \ell b \bar{b}$ events before double b-tagging.
$\varepsilon _{b}$ is the efficiency of double b-tagging for $Zbb$ signal events which is determined from
                           the Monte-Carlo simulation tuned to reproduce the b-tagging efficiency measured from the data, 
$\varepsilon _{c}$  is the efficiency of double b-tagging for $Zcc$+jets background events which is determined 
                           from the Monte Carlo,
$\varepsilon _{\ell}$ is the efficiency of double b-tagging for $Z$+jets (jet=u,d,s,g) background events which 
                              is determined from the Monte-Carlo simulation tuned to reproduce the mistagging efficiency measured 
                              from the data.

The number of unknown parameters are reduced to two, following the D0 analysis approach \cite{Abazov:2004zd}, by using the theoretical ratio of cross sections and ratio of selection efficiencies
\begin{equation}
 R~=~\frac{\sigma (Zcc)}{\sigma (Zjj)}~\times~\frac{\varepsilon_{Zcc}^{sel}}{\varepsilon_{Zjj}^{sel}}~=~0.046~\pm~0.002  \\
\end{equation}
and by replacing $N_{Zcc}$ by $R \times N_{Zjj}$. The uncertainty of the $\frac{\sigma (Zcc)}{\sigma (Zjj)}$ ratio due to $\mu_{R}$, $\mu_{F}$ scale variation and JES and MET scale are considered. The equations are solved to evaluate the  value of $N_{Zbb}$. The uncertainty on $N_{Zbb}$, $\Delta N_{Zbb}$,  is than calculated due to the uncertainties of $N_{Z}^{after~b-tag}$ from 
$t\bar{t}$ background subtraction ($\delta N_{Z}^{after~b-tag}$=4.0\%), the uncertainty of $R$ and the uncertainties of $\varepsilon _{b}$, $\varepsilon _{c}$ and $\varepsilon _{\ell}$. From the estimated number of $N_{Zbb}$ events after all selections except b-tagging, the number of $Zbb$ events before the lepton and jet selections and cut on $E_{T}^{miss}$ can be evaluated. The systematic uncertainties due to Jet Energy Scale, $E_{T}^{miss}$ scale, lepton selections, b-tagging, Monte Carlo dependence on $p_{T}$ and $\eta$ of jets and luminosity measurement have been taken into account according to the values foreseen to be achieved in CMS with data corresponding to 100 pb$^{-1}$ of integrated luminosity. 

The statistical uncertainty is defined as $\Delta N_{sel}/N_{sel}$=$1/\surd N_{sel}$, where $N_{sel}$ is the measured number of events after all selections (46), thus it is $\delta N_{sel}$=14.7\%. 

Finally, the $\ell \ell b \bar{b}$ ($\ell$=e, $\mu$) cross-section is expected to be measured with an accuracy of

\begin{equation}
\delta \sigma~=~\pm~^{21\%}_{25\%}~(syst)~\pm~15\%~(stat.)
\end{equation}

\section{Conclusion}

The process $pp\rightarrow b\bar{b}Z$ at LHC needs to be well understood  because of its importance as a background to SM and SUSY Higgs boson discovery channels. The possibility of measuring the $b\bar{b}Z, Z\rightarrow \ell\ell$ process at CMS with the first 100~pb$^{-1}$ of 
data has been studied with a robust selection of leptons and b-jets. Possible methods of measuring backgrounds 
from data have been discussed. The statistical and systematic uncertainties have been discussed.

\end{document}